\documentclass[useAMS,usenatbib,fleqn]{mn2e}

\usepackage{amsmath,amssymb}
\usepackage{bm}
\usepackage{graphicx}

\newcommand{\aj}{AJ}
\newcommand{\apj}{ApJ}
\newcommand{\apjl}{ApJ}
\newcommand{\apjs}{ApJS}
\newcommand{\araa}{ARA\&A}
\newcommand{\mnras}{MNRAS}
\newcommand{\newa}{NewA}
\newcommand{\nat}{Nature}
\newcommand{\science}{Science}
\newcommand{\pasj}{PASJ}

\usepackage{ulem}
\usepackage{color}

\title[Merger of Multiple Massive Black Holes]{Merger Criteria of
  Multiple Massive Black Holes and the Impact on the Host Galaxy}
\author[A. Tanikawa and
  M. Umemura]{A. Tanikawa$^{1,2,3}$\thanks{E-mail:
    ataru.tanikawa@riken.jp} and M. Umemura$^{1}$\\ $^{1}$Center for
  Computational Sciences, University of Tsukuba, 1-1-1, Ten-nodai,
  Tsukuba, Ibaraki 305-8577, Japan\\ $^{2}$School of Computer Science
  and Engineering, University of Aizu, Tsuruga, Ikki-machi,
  Aizu-Wakamatsu, Fukushima 965-8580, Japan\\ $^3$RIKEN Advanced
  Instistute for Computational Science, 7-1-26,
  Minatojima-minami-machi, Chuo-ku, Kobe, Hyogo, 650-0047, Japan}
\begin{document}

\date{Accepted 1988 December 15. Received 1988 December 14; in original form 1988 October 11}

\pagerange{\pageref{firstpage}--\pageref{lastpage}} \pubyear{2002}

\maketitle

\label{firstpage}

\begin{abstract}
  We perform $N$-body simulations on a multiple massive black hole
  (MBH) system in a host galaxy to derive the criteria for successive
  MBH merger. The calculations incorporate the dynamical friction by
  stars and general relativistic effects as pericentre shift and
  gravitational wave recoil.  The orbits of MBHs are pursed down to
  ten Schwarzschild radii ($\sim 1$AU).  As a result, it is shown that
  about a half of MBHs merge during 1~Gyr in a galaxy with mass
  $10^{11}M_{\odot}$ and stellar velocity dispersion
  $240$~km~s$^{-1}$, even if the recoil velocity is two times as high
  as the stellar velocity dispersion. The dynamical friction allows a
  binary MBH to interact frequently with other MBHs, and then the
  decay of the binary orbits leads to the merger through gravitational
  wave radiation, as shown by \citet{Tanikawa11}. We derive the MBH
  merger criteria for the masses, sizes, and luminosities of host
  galaxies. It is found that the successive MBH mergers are expected
  in bright galaxies, depending on redshifts.  Furthermore, we find
  that the central stellar density is reduced by the sling-shot
  mechanism and that high-velocity stars with $\sim 1000$~km~s$^{-1}$
  are generated intermittently in extremely radial orbits.
\end{abstract}


\section{Introduction}

Massive black holes (hereafter MBHs) with the mass of more than $10^6$
solar mass ($M_{\odot}$) have been found in the centres of galaxies.
The mass of MBHs is correlated with the properties of the spheroidal
components of their host galaxies, with respect to the mass
\citep{Kormendy95,Magorrian98,Marconi03}, the velocity dispersions
\citep{Ferrarese00,Tremaine02,Gultekin09}, and the number of globular
clusters \citep{Burkert10,Harris11}. The origin of MBHs is an open
issue of great significance.

In the last decade, quasars (QSOs) that possess $\sim 10^9~M_{\odot}$
MBHs have been found at high redshifts of $z \gtrsim 6$
\citep[e.g.,][]{Fan01}, that is, at the cosmic age of $\lesssim
1$~Gyr. Conservatively speaking, the seeds of the MBHs could be
stellar mass black holes as massive star remnants. In particular, the
remnants of first stars are one of plausible candidates, since first
stars are likely to be as massive as a few hundred solar mass
\citep{Abel00,Nakamura01,Bromm02,Yoshida06}, several tens solar mass
\citep{Clark11}, or about $50M_{\odot}$ \citep{Hosokawa11}. However,
in order for first star remnants to grow up to $\sim 10^9~M_{\odot}$
in $1$~Gyr, the Eddington ratio of mass accretion rate should be
larger than unity \citep[e.g.,][and references
  therein]{Umemura01,Greene12}. Super-Eddington accretion is one of
possible solutions for the MBH growth
\citep[e.g.][]{Abramowicz88,Kawaguchi03,Ohsuga05}.  On the other hand,
the integration of the QSO luminosity function is concordant with the
integrated mass function of MBHs in the local universe, as long as the
Eddington ratios are between 0.1 and 1.7
\citep{Soltan82,Yu02,Marconi04}. This implies that supermassive black
holes acquire the bulk of mass through gas accretion in the late
evolutionary stages and the mass accretion rates are not highly
super-Eddington. Also, the gas accretion onto the seeds should be
intermittent, and on average could be lower than the Eddington
accretion rate \citep{Milosavljevic09a,Milosavljevic09b}.  If the
merger of multiple black holes precedes the growth via gas accretion,
the merged MBH can be a seed of a supermassive black hole, and
therefore the constraint for the BH growth can be weaker.

In the cold dark matter cosmology, larger galaxies form hierarchically
through mergers of smaller galaxies. Hence, many MBHs are assembled in
a larger galaxy, if smaller galaxies already possess MBHs.
Furthermore, MBHs could be born in hyper-massive star clusters formed
by galaxy collisions \citep{Matsui11}. Thus, galaxy merger
remnant can contain many MBHs, even if precursory galaxies have no
MBHs.

Observationally, multiple active galactic nucleus (AGN) systems have
been discovered recently.  They include a triple AGN in the galaxy
SDSS J1027+1749 at $z = 0.066$ \citep{Liu11}, three rapidly growing
MBHs of $10^6-10^7 M_\odot$ in a clumpy galaxy at $z = 1.35$
\citep{Schawinski11}, a firstly-discovered physical quasar triplet QQQ
J1432--0106 within the projected separation of $30-50$kpc at $z =
2.076$ \citep{Djorgovski07}, and a second quasar triplet QQQ
J1519+0627 within the projected separation of $200$kpc, which is
likely to be harboured in a yet-to-be-formed massive system at $z=1.51$
\citep{Farina13}.  According to the hierarchical merger history,
galaxies with many MBHs are likely to form at higher redshifts.
Although the galaxy merger proceeds through the violent relaxation,
the merger of MBHs has difficulty.  As pointed out in
\citet{Begelman80}, two MBHs in a galaxy are likely to form a binary,
but unlikely to merge directly due to the so-called loss cone
depletion (the depletion of stars on orbits that intersect the binary
MBH).  A binary MBH cannot reach sub-parsec separation due to the loss
cone depletion, which is called the final parsec problem
\citep[e.g.][]{Merritt04}.  A possible way to evade the loss cone
depletion is the nonaxisymmetric potential of the host galaxy
\citep{Merritt04,Berczik06,Khan11,Khan12}, which is the natural
consequence of the galaxy merger.  A binary MBHs can merge in the
nonaxisymmetric potential in $10$~Gyr or $0.3$~Gyr, when the galaxy
contains stars with $10^9M_{\odot}$ or $10^{11}M_{\odot}$,
respectively \citep{Khan11}.  However, since this timescale is
comparable to or longer than the galactic dynamical timescale, other
galaxies harbouring MBHs can intrude before two MBHs merge.  This is
likely to occur at higher redshifts of $z \gtrsim 6$, at which the
universe age is less than $1$~Gyr. If there are more than two MBHs in
a galaxy, the dynamical relaxation of MBHs is significantly controlled
by the gravity of MBHs themselves, especially by three-body
interaction.  When a third MBH intrudes into a binary MBH, one of the
three MBHs carries away angular momentum from the rest two MBHs,
reducing the binary separation, and eventually the binary merges
\citep[e.g.][]{Iwasawa06}.

So far, galaxy structures have not been investigated when the galaxies
contain more than three MBHs, although they have been investigated in
the cases of two MBHs \citep[e.g.][]{Khan11,Khan12} and three MBHs
\citep[e.g.][]{Iwasawa08}.  \cite{Tanikawa11} (hereafter, paper~I)
scrutinised a system of multiple MBHs in a galaxy by high-resolution
$N$-body simulations, and found that multiple MBHs produce one
dominant MBH through successive mergers. Binary MBHs lose their
angular momentum owing to sling-shot mechanism, which induces the
decay of the binary orbits through gravitational wave (GW)
radiation. In paper~I, we investigated one model of a galaxy
containing multiple MBHs. In this paper, we explore the evolution of
multiple MBHs in galaxies with different 3-dimensional stellar
velocity dispersions to derive the criteria of the MBH merger. We also
consider the effect of the recoil by anisotropic GW radiation at the
MBH merger.  Since the recoil velocity typically reaches several
hundred km~s$^{-1}$ \citep{Kesden10}, it could suppress the MBH
growth. Furthermore, we investigate the impact by the MBH merger on
the galaxy structure.

The paper is organised as follows. In section \ref{sec:model}, we
describe the simulation model. In section \ref{sec:results}, we show
numerical results. In section \ref{sec:criteria}, the results are
translated to derive the criteria for MBH merger, which are applied
for high and low redshift galaxies.  In section
\ref{sec:back-reaction}, the back-reaction to a host galaxy is
discussed with respect to the galactic structure and the production of
high velocity stars.  In section \ref{sec:summary}, we summarise this
paper.

\section{Model}
\label{sec:model}

\subsection{Initial conditions}

We consider a model galaxy that initially contains ten MBHs of equal
mass. The effect by the inequality of MBH mass has been explored by
several authors \citep[e.g.][]{Iwasawa11,Khan12}.  In the present
simulation, an unequal mass binary forms as a consequence of the MBH
merger.  The case in which unequal mass MBHs are set up initially will
be investigated elsewhere. Stars in a galaxy are treated as
superparticles.  The number of stars is $N=512k$ ($1k=1024$).  The
stars are initially distributed according to the Hernquist's profile,
where the mass density distribution is given by
\begin{equation}
  \rho (r) = \frac{M_{\rm g}}{6 \pi r_{\rm g}^3} \frac{1}{(r/r_{\rm
      g})\left[ (r/r_{\rm g}) + 1/3 \right]^3},
\end{equation}
where $M_{\rm g}$ and $r_{\rm g}$ are respectively the total mass and
virial radius of the host galaxy. Here, $r_{\rm g}$ is given by
\begin{equation}
  r_{\rm g} = \frac{GM_{\rm g}}{2v_{\rm g}^2}, \label{eq:virial}
\end{equation}
with the gravitational constant $G$ and the 3-dimensional stellar
velocity dispersion $v_{\rm g}$.

The mass of MBH is set to be $0.01$ \% of the galaxy mass. So, the
total mass of ten MBHs is $0.1$ \% of the galaxy mass. We realise the
distribution of MBHs as follows. The distribution function is supposed
to be the same as that of stars within one-third of $r_{\rm g}$ (see
Paper~I for the dependence on the spread). First, we generate
positions and velocities for stars according to the above
distributions. Next, we convert ten stars into ten MBHs; we choose
randomly ten stars from the stars within one-third of $r_{\rm g}$.

The 3-dimensional velocity dispersion, $v_{\rm g}$, is one of key
parameters in the present simulations. Note that the velocity
dispersion in this paper is 3-dimensional unless otherwise noted. We
consider several galaxy models with different velocity
dispersions. The assumed models are shown in
Table~\ref{tab:mainresult}. In models A, the velocity dispersion is
$v_{\rm g}=350$~km~s$^{-1}$, where A$_{0,1}$, A$_{0,2}$, and A$_{0,3}$
are based on different sets of random numbers. In models B, C, and D,
$v_{\rm g}=240$, 180, and 120~km~s$^{-1}$, respectively. In models
A$_1$, B$_1$, and C$_1$, the recoil velocity, $v_{\rm recoil}$, is
added after the MBH merger. Also, for the comparison to models with
ten MBHs, we perform simulations for galaxies without MBHs (model BH0)
and with two MBHs (model BH2).

In the present simulations, we adopt the standard $N$-body units,
where $G=M_{\rm g}=r_{\rm g}=1$. Then, $v_{\rm g}=1/\sqrt{2}$ from
Equation~(\ref{eq:virial}).  The speed of light, $c$, is required to
be redefined in the present units, since we include Post-Newtonian
(PN) corrections (described later).  The speed of light changes as
$c=6.06 \times 10^2$, $8.84 \times 10^2$, $1.18 \times 10^3$, or $1.77
\times 10^3$ for models A, B, C, or D, respectively.

After we determine the velocity dispersion, $v_{\rm g}$, we still have
one free parameter, although $M_{\rm g}/r_{\rm g}$ is fixed for each
$v_{\rm g}$ (see Equation~(\ref{eq:virial})). Setting either of the
galaxy mass $M_{\rm g}$ or the galactic virial radius $r_{\rm g}$, we
can transform the code units to physical units. When we set $M_{\rm
  g}$, we express $r_{\rm g}$ and the dynamical time at $r_{\rm g}$ as
follows:
\begin{equation}
  r_{\rm g} \simeq 1.76 \left( \frac{M_{\rm g}}{10^{11}~M_{\odot}}
  \right) \left( \frac{v_{\rm g}}{350~\mbox{km~s$^{-1}$}} \right)^{-2}
  \mbox{[kpc]}, \label{eq:rv}
\end{equation}
and 
\begin{equation}
  t_{\rm dy,g} = \frac{r_{\rm g}}{\sqrt{2}v_{\rm g}} \simeq 3.47
  \left( \frac{M_{\rm g}}{10^{11}~M_{\odot}} \right) \left(
  \frac{v_{\rm g}}{350~\mbox{km~s$^{-1}$}} \right)^{-3}
  \mbox{[Myr]}. \label{eq:tdy}
\end{equation}
We show $M_{\rm g}$ when $r_{\rm g}=1$~kpc in the rightmost column of
Table~\ref{tab:mainresult}, which is derived from
Equation~(\ref{eq:rv}). The average mass density inside $r_{\rm g}$ is
given as
\begin{equation}
  \rho_{\rm g} = \frac{27M_{\rm g}}{64\pi r_{\rm g}^3} \simeq 2.48
  \left( \frac{M_{\rm g}}{10^{11} M_{\odot}} \right)^{-2} \left(
  \frac{v_{\rm g}}{350~\mbox{km~s$^{-1}$}} \right)^6
  \mbox{[$M_\odot~\mbox{pc}^{-3}$]}.
\end{equation}

\subsection{Equation of motion}

The equations of motion for field stars and MBHs are respectively
given by
\begin{align}
  \frac{d^2\bm{r}_{{\rm f},i}}{dt^2} &= \sum_{j \neq i}^{N_{\rm
      f}} \bm{a}_{{\rm ff},ij} + \sum_{j}^{N_{\rm B}}
  \bm{a}_{{\rm fB},ij}, \\ 
\frac{d^2\bm{r}_{{\rm B},i}}{dt^2}
  &= \sum_{j}^{N_{\rm f}} \bm{a}_{{\rm Bf},ij} + \sum_{j \neq
    i}^{N_{\rm B}} \bm{a}_{{\rm BB},ij},
\end{align}
where $\bm{r}_{{\rm f},i}$ and $\bm{r}_{{\rm B},i}$ are the
position vectors of $i$-th field star and $i$-th MBH, $N_{\rm f}$ and
$N_{\rm B}$ are the numbers of field stars and MBHs, $\bm{a}_{{\rm
    ff},ij}$ and $\bm{a}_{{\rm fB},ij}$ are the accelerations by
$j$-th field star and $j$-th MBH on $i$-th field star, and
$\bm{a}_{{\rm Bf},ij}$ and $\bm{a}_{{\rm BB},ij}$ are the
accelerations by $j$-th field star and $j$-th MBH on $i$-th MBH,
respectively.  Excepting the MBH-MBH interaction, the accelerations
are given by Newtonian gravity:
\begin{align}
  \bm{a}_{{\rm ff},ij} &= - Gm_{{\rm f},j}\frac{\bm{r}_{{\rm
        f},i} - \bm{r}_{{\rm f},j}}{(|\bm{r}_{{\rm f},i} -
    \bm{r}_{{\rm f},j}|^2+\epsilon^2)^{3/2}} \\ \bm{a}_{{\rm
      fB},ij} &= - Gm_{{\rm B},j}\frac{\bm{r}_{{\rm f},i} -
    \bm{r}_{{\rm B},j}}{|\bm{r}_{{\rm f},i} - \bm{r}_{{\rm
        B},j}|^3} \\ \bm{a}_{{\rm Bf},ij} &= - Gm_{{\rm
      f},j}\frac{\bm{r}_{{\rm B},i} - \bm{r}_{{\rm
        f},j}}{|\bm{r}_{{\rm B},i} - \bm{r}_{{\rm f},j}|^3},
\end{align}
where $m_{{\rm f},j}$ and $m_{{\rm B},j}$ are respectively the masses
of $j$-th field star and $j$-th MBH, and the softening parameter
($\epsilon=10^{-3}$) is introduced only in star-star interactions.

The acceleration between two MBHs is composed of the Newtonian gravity
and PN corrections, such as
\begin{equation}
  \bm{a}_{{\rm BB},ij} = - Gm_{{\rm B},j}\frac{\bm{r}_{{\rm
        B},i}-\bm{r}_{{\rm B},j}}{|\bm{r}_{{\rm
        B},i}-\bm{r}_{{\rm B},j}|^3} + \bm{a}_{{\rm
      PN},ij}. \label{eq:aBB}
\end{equation}
We explain the second term below.

\subsection{Relativistic effects}

We incorporate the general relativistic effects on the orbits of MBHs,
that is, the pericentre shift, GW radiation, and GW recoil.  We model
the pericentre shift and GW radiation by including the second term
($\bm{a}_{{\rm PN},ij}$) in Equation~(\ref{eq:aBB}) up to 2.5PN
term.  The pericentre shift corresponds to 1PN and 2PN terms, and the
GW radiation does to 2.5PN term \citep{Damour81,Soffel89,Kupi06}.  We
employ Equations (1), (2), (3), and (4) in \cite{Kupi06} for the
general relativistic corrections.

Also, we model the GW recoil as follows. At the moment when two MBHs
merge, we add recoil velocities to the merged MBHs. Their absolute
values are fixed in each simulation. Their direction is determined by
the Monte-Carlo method, assuming the isotropic probability.  In
practice, the absolute values of recoil velocities widely range from
several ten km~s$^{-1}$ to several thousand km~s$^{-1}$, and their
directions are determined by the mass ratio and spins of two MBHs
\citep{Campanelli07,Lousto10}. However, if their spins are aligned
before their merger due to relativistic spin precession
\citep{Kesden10}, then the recoil velocity decreases to a few $100$
km~s$^{-1}$.  The recoil velocity $v_{\rm recoil}$ in each simulation
is summarised in Table~\ref{tab:mainresult}. We set the recoil
velocity to be equal to or more than $200$~km~s$^{-1}$, excepting
models without the recoil.

\subsection{Merger condition}

We assume that two MBHs merge, when the separation between two MBHs is
less than ten times the sum of their Schwarzschild radii:
\begin{equation}
  \left|\bm{r}_{{\rm B},i} - \bm{r}_{{\rm B},j} \right| < 10
  \left( r_{{\rm sch},i} + r_{{\rm sch},j} \right),
\end{equation}
where $r_{{\rm sch},i}$ is the Schwarzschild radius of $i$-th MBH
that is $2Gm_{{\rm B},i}/c^2$ for
the MBH mass $m_{{\rm B},i}$ with the speed of light $c$ .

\subsection{Numerical scheme}

Since our numerical scheme is the same as that in paper~I, we describe
its outline here. We adopt a fourth-order Hermite scheme with
individual timestep scheme \citep{Makino92} for time integration
method for an MBH and a star.

For compact binary MBHs, we transform their motions to their relative
motion and the centre-of-mass motion. The relative motions are
integrated in the same way as single MBHs and stars. In calculating
tidal forces on the binary MBHs, we consider the other MBHs and nearby
stars as perturbers, and ignore perturbation by the rest of stars.  A
perturber of a binary MBH is defined as a particle whose distance from
the binary MBH is smaller than $200$ times of the semi-major axis of
the binary MBH.  For the centre-of-mass motion, we adopt Hermite
Ahmad-Cohen scheme \citep{Makino92} for time integration.

We perform $N$-body simulations with the FIRST simulator
\citep{Umemura08} at University of Tsukuba. We use 64 nodes of the
FIRST simulator. Each node is equipped with one Blade-GRAPE, which is
one of GRAPEs: a special purposed accelerator for a collisional
$N$-body system \citep{Sugimoto90,Makino03,Fukushige05}. We compute
gravitational forces exerting on a given particle in parallel, which
is the so-called $j$-parallel algorithm.

\section{Numerical Results}
\label{sec:results}

\subsection{Model dependence}

We have calculated a system of ten MBHs in one galaxy during about
$140\, t_{\rm dy,g}$, which corresponds to about 1~Gyr in physical
units if we adopt $M_{\rm g}=10^{11} M_{\odot}$ and $v_{\rm
  g}=240$~km~s$^{-1}$. The models and results are summarised in
Table~\ref{tab:mainresult}. The first and second columns indicate the
model name and the number of MBHs, respectively. Models A$_{0,1}$,
A$_{0,2}$, and A$_{0,3}$ corresponds to models A$_1$, A$_{2}$, and
A$_{3}$ in Paper~I, respectively. The third column is the stellar
velocity dispersion. In the fourth and fifth columns, we show the
assumed GW recoil velocity and the ratio of the recoil velocity to the
velocity dispersion, respectively. As numerical results, we show the
mass of the heaviest MBH ($m_{\rm B,p}$) and the second heaviest MBH
($m_{\rm B,s}$) in the sixth and seventh columns, respectively. Their
mass is scaled by the initial MBH mass. If the heaviest or second
heaviest MBHs are ejected from the galaxy centre, we attach ``(e)''
beside their mass. The number of MBHs ejected from the galaxy centre
($N_{\rm B,ej}$) is shown in the eighth column. We define the ejected
MBHs to be far by more than $r_{\rm g}$ from the galaxy centre. The
galaxy centre is obtained from the density centre of stars, which is
calculated in the same way as \cite{Casertano85}.

\begin{figure}
  \begin{center}
    \includegraphics[scale=1]{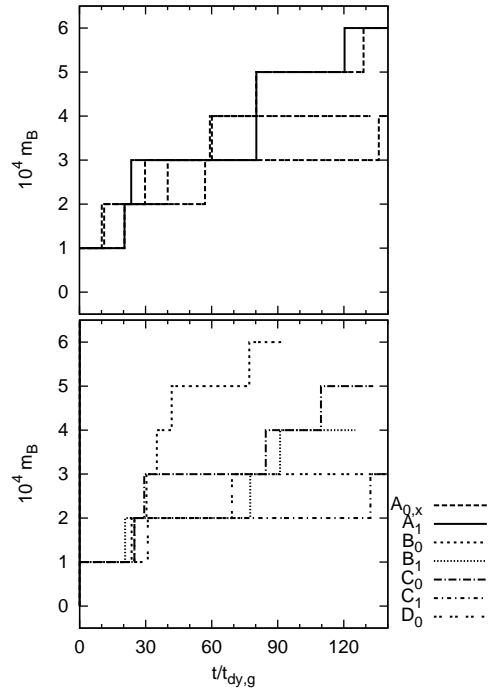}
  \end{center}
  \caption{Time evolution of the mass of the heaviest MBH in each
    model. Models A$_{0,1}$, A$_{0,2}$, and A$_{0,3}$ are brought
    together as ``A$_{0,x}$''. Models A are shown in the top panel,
    and the other in the bottom panel.}
  \label{fig:mmax}
\end{figure}

Fig.~\ref{fig:mmax} shows the time evolution of the mass of the
heaviest MBH in each model.  In each of model A, B, and C$_0$, one
dominant MBH grows in a galaxy. They are formed through mergers of $4$
-- $6$ MBHs. In some of these models, other MBHs become heavier than
the initial ones. However, they are ejected from the galaxy through
sling-shot mechanism by three MBH interaction.  They are not ejected
by the GW recoil, since the GW recoil is set to be at most $2v_{\rm
  g}$, while the escape velocity of our galaxy models is about
$4v_{\rm g}$ (described below in detail).  In these models, about a
half of MBHs successively merge in $140\, t_{\rm dy,g}$.

On the other hand, in models C$_1$ and D$_0$, only three MBHs merge in
$140\, t_{\rm dy,g}$.  A heavier MBH might form, if we follow the
evolution of the MBHs beyond $140\, t_{\rm dy,g}$. However, we do not
follow their evolution, since artificial two-body relaxation may
affect the merger for the present number of particles.

\begin{table*}
  \caption{MBH mass and the number of ejected MBHs after $140 t_{\rm
      dy,g}$. The units of $m_{\rm B,p}$ and $m_{\rm B,s}$ are initial
    MBH mass.}
  \label{tab:mainresult}
  \begin{center}
    \begin{tabular}{c|c|c|cc|c|c|c|c}
      \hline
      Model & $N_{B}$ &
      $v_{\rm g}/\mbox{km s}^{-1}$ &
      ${v_{\rm recoil}}/\mbox{km s}^{-1}$
      & $v_{\rm recoil}/v_{\rm g}$ &
      $m_{\rm B,p}$ &
      $m_{\rm B,s}$
      & $N_{\rm B,ej}$
      & $M_{\rm g}/(10^{10}M_{\odot})$ $[r_{\rm g}=\mbox{1kpc}]$ \\
      \hline
      \hline
      A$_{0,1}$ & $10$ & $350$ & $0$    & $0$   & $4$ & $3$(e) & $2$ & $5.7$ \\
      A$_{0,2}$ & $10$ & $350$ & $0$    & $0$   & $4$ & $1$    & $3$ & $5.7$ \\
      A$_{0,3}$ & $10$ & $350$ & $0$    & $0$   & $6$ & $1$    & $1$ & $5.7$ \\
      A$_1$     & $10$ & $350$ & $500$  & $1.4$ & $6$ & $1$    & $1$ & $5.7$ \\
      \hline
      B$_0$     & $10$ & $240$ & $0$    & $0$   & $6$ & $1$    & $1$ & $2.7$ \\
      B$_1$     & $10$ & $240$ & $500$  & $2.1$ & $4$ & $2$(e) & $2$ & $2.7$ \\
      \hline
      C$_0$     & $10$ & $180$ & $0$    & $0$   & $5$ & $1$    & $1$ & $1.5$ \\
      C$_1$     & $10$ & $180$ & $200$  & $1.1$ & $3$ & $2$    & $1$ & $1.5$ \\
      \hline
      D$_0$     & $10$ & $120$ & $0$    & $0$   & $3$ & $1$    & $2$ & $0.67$ \\
      \hline
      BH0       & $0$ & $240$    & --     & --    & --  & --   & --  & $2.7$ \\
      BH2       & $2$ & $240$    & --     & --    & --  & --   & --  & $2.7$ \\
      \hline
    \end{tabular}
  \end{center}
\end{table*}

\subsection{Merger dynamics}

\begin{figure}
  \begin{center}
    \includegraphics[scale=1]{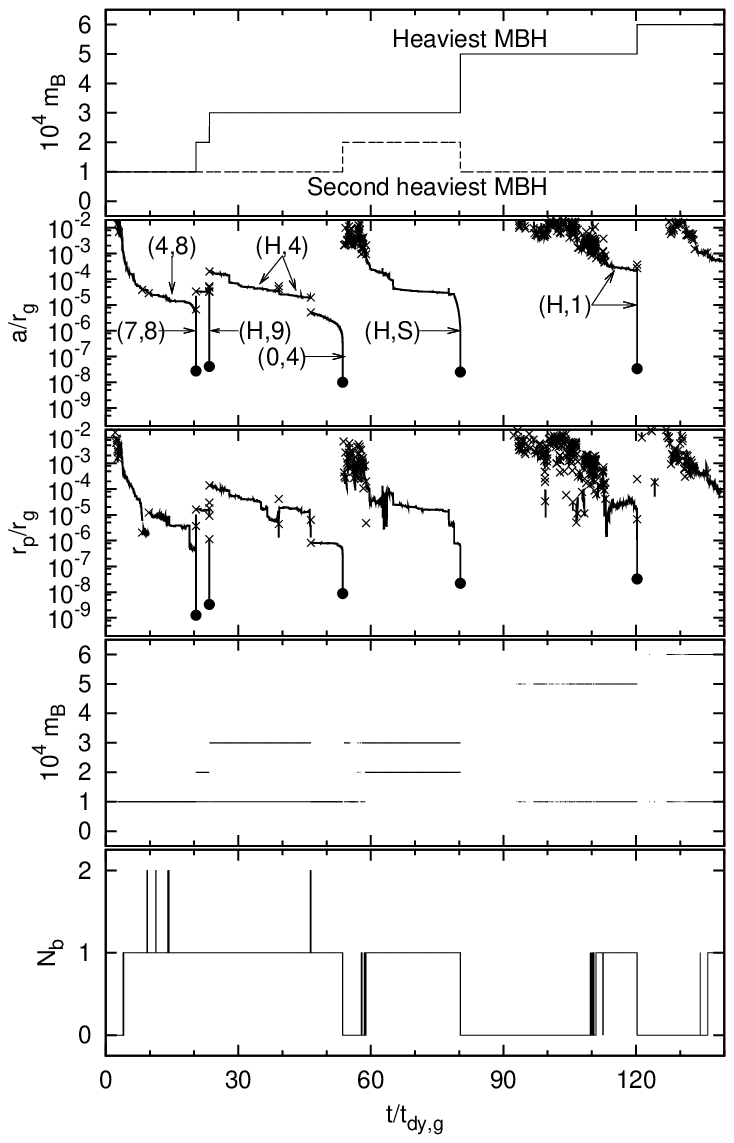}
  \end{center}
  \caption{Time evolution of MBHs in model A$_1$. The top panel shows
    the masses of the heaviest and second heaviest MBHs at each
    time. We indicate parameters of the binary MBH with the smallest
    semi-major axis at each time, i.e. its semi-major axis (second
    top), the distance at the pericentre (middle), and component
    masses (second bottom). The bottom panel shows the number of
    binary MBHs whose semi-major axes are less than $10^{-3}r_{\rm
      g}$. Pairs of integers in parentheses in the second top panel
    show the labels of MBHs composing the binary MBHs, where the
    heaviest MBH is labelled with "H" and the second heaviest one is
    "S". We attached labels only to binary MBHs which are long-lived,
    or merge eventually. In the second top and middle panels, filled
    circles indicate the moments when MBHs merge and crosses denote
    those when binary components are exchanged. }
  \label{fig:bbh}
\end{figure}

Here, we see the merger process in detail, using the result of model
A$_1$, which includes the GW recoil. The merger process is similar to
that in the models without the GW recoil, i.e., models A$_{0,1}$,
A$_{0,2}$, and A$_{0,3}$, in which the merger process is shown in
Paper~I. As seen in the top panel of Fig.~\ref{fig:bbh}, one MBH grows
by the merger, and other MBHs do not grow by the terminal time of our
simulation. Two MBHs temporarily merge at $t_{\rm dy,g} =54$.
However, the merged MBH is swallowed by the heaviest MBH at $t_{\rm
  dy,g} = 80$.

The result that only one MBH predominantly grows comes from the
following three facts. Two MBHs merge only via a phase of a binary MBH
(fact~1). The binary MBH tends to contain the heaviest MBH
(fact~2). Furthermore, the binary MBH is unique in the galaxy at any
time (fact~3). Fact~1 can be verified in the second top panel of
Fig.~\ref{fig:bbh}. Binary MBHs have semi-major axes of $10^{-5}r_{\rm
  g}$ -- $10^{-4}r_{\rm g}$ for a long time until they merge. Fact~2
can be seen in the second bottom panel. The heaviest MBH is contained
in a binary MBH through most of time (except during $t_{\rm dy,g}=46$
-- $54$). This is because a binary MBH often experiences three MBH
interactions, through which a heavier MBH is more easily retained in
the binary MBH. Fact~3 can be confirmed in the bottom panel. For most
of time, the number of binary MBHs in the galaxy is one or zero.

For the mergers of MBHs, the dynamical friction plays a key role.
(see also Fig. 2 in Paper I). The dynamical friction by field stars
allows MBHs to gather near the galaxy centre. Thus, two MBHs can
compose a binary MBH, and subsequently another MBH can intrude the
binary MBH. Then, the single MBHs interact with the binary MBH
repeatedly and consequently the semi-major axis and eccentricity of
the binary MBH are changed owing to the angular momentum loss. Such
repeated interactions occur before most of mergers. However, the
crucial impact is brought by one strong interaction, and thereby the
distance of the binary MBH at the pericentre ($r_{\rm p}$) shrinks
significantly, so that the GW radiation works effectively to lose the
energy, eventually causing the merger of the binary.

Such merger process is not affected by the GW recoil, if the recoil
velocity is of the order of the stellar velocity dispersion, $v_{\rm
  g}$. A merged MBH is retained in the inner region of the galaxy, in
which the stellar density is high. In this region, the dynamical
friction effectively loses angular momenta of the merged MBH. Hence,
the merged MBH falls again toward the galactic centre, and form a
binary MBH with another single MBH. The binary MBH can interact again
a third MBH.

Also, we have found that the secular angular momentum loss of a binary
MBH through the Kozai mechanism \citep{Kozai62} is not effective.  The
Kozai mechanism can work through eccentricity oscillation, if the
semi-major axis ratio is small \citep{Blaes02,Berentzen09}.  But, in
our simulations, the semi-major axis ratio is too large to allow
eccentricity oscillation. Instead, the relativistic pericentre shift
(1PN and 2PN) is dominant.  In fact, if we do not include the 1PN and
2PN terms, the Kozai mechanism works for a binary to merge.  The
suppression of the Kozai mechanism is also demonstrated in the case of
stellar-sized black holes \citep{Miller02} and in the planetary orbits
\citep{Fabrycky07}.

\section{Criteria for Successive Mergers}
\label{sec:criteria}

\subsection{Constraints for galaxy mass and size}

\begin{figure*}
  \begin{center}
    \includegraphics[scale=1.5]{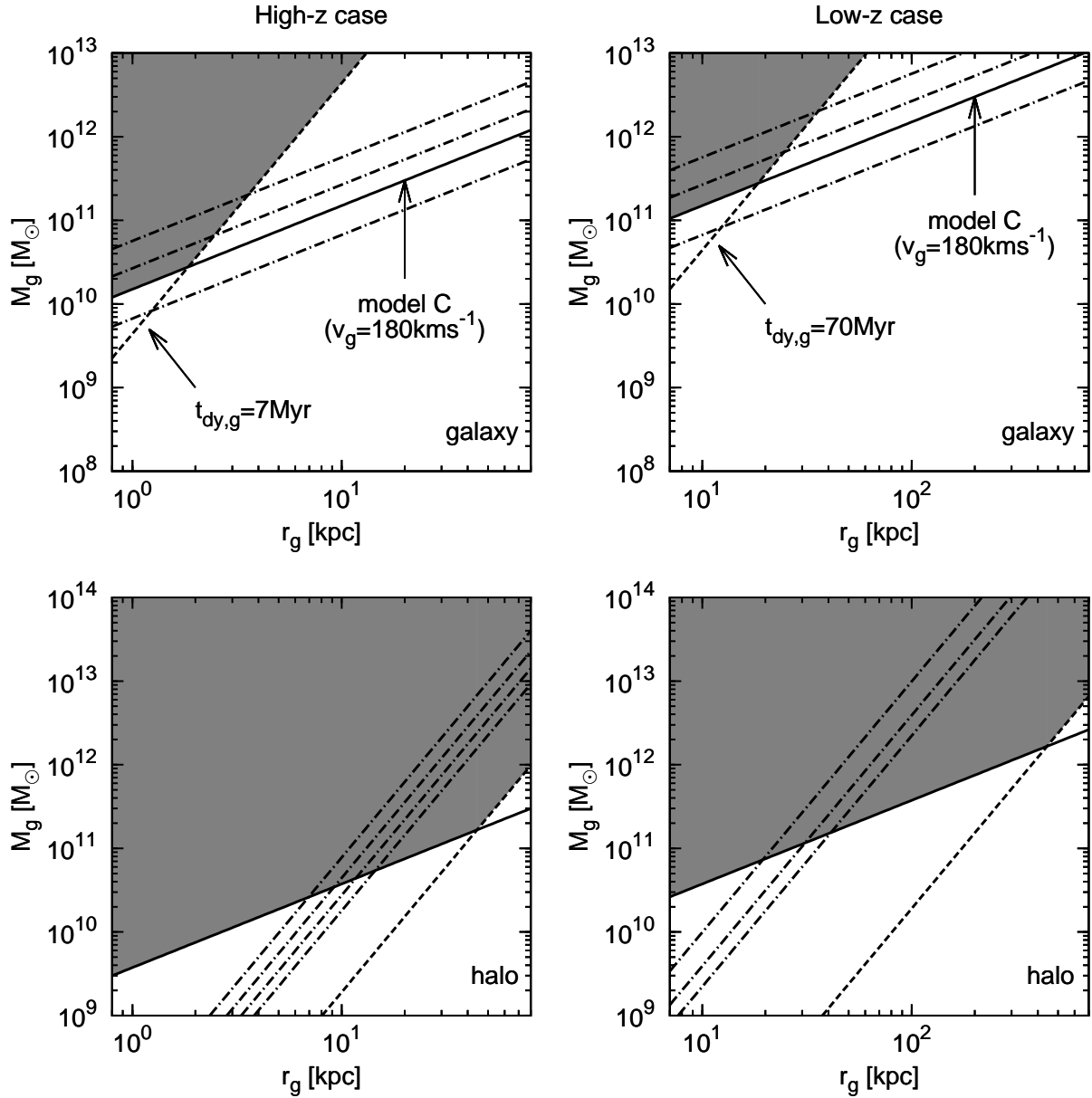}
  \end{center}
  \caption{Mass and size of a galaxy (top panels) and halo containing
    the galaxy (bottom panels). In left and right panels, gray regions
    indicate the galaxy and halo in which MBHs merge within $1$~Gyr
    and $10$~Gyr, respectively. In the top panels, the solid lines
    correspond to a galaxy of model C, and the dashed lines indicate
    galaxies with $t_{\rm dy,g}=7$~Myr (left) and $70$~Myr
    (right). The dashed-dotted lines in the top left and right panels
    correspond to galaxies of models A, B, and D from top to
    bottom. In the bottom panels, the dashed-dotted lines show mass
    and size of a halo formed at redshift $z=15$, $10$, $7$, and $5$
    (from left to right) in the left panel, and those of a halo formed
    at redshift $z=3$, $1$, and $0$ (from left to right) in the right
    panel.}
  \label{fig:lmt}
\end{figure*}

\begin{figure*}
  \begin{center}
    \includegraphics[scale=1.5]{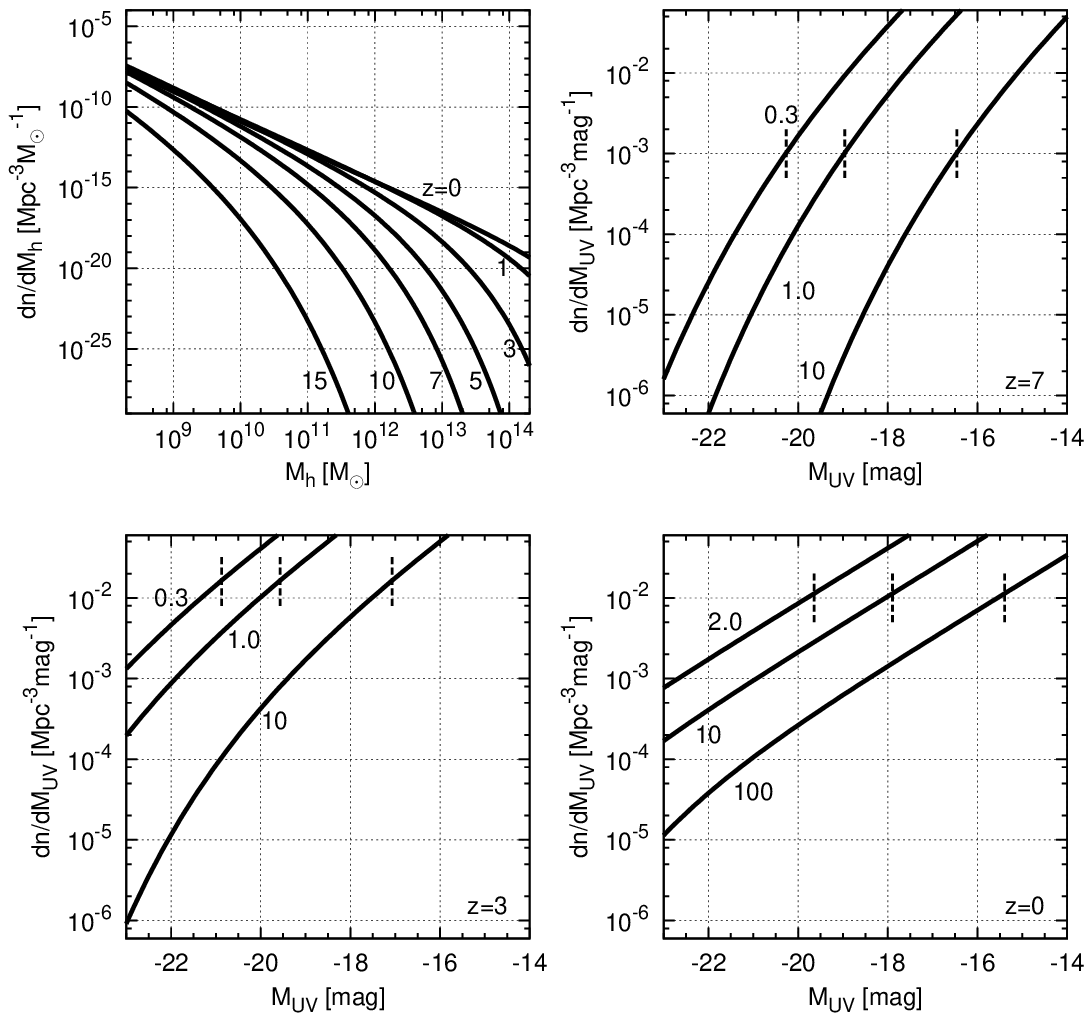}
  \end{center}
  \caption{(Top left) Press-Schechter mass function at a given
    redshift. Cosmological constants are set as $(\Omega_{\rm
      \Lambda}, \Omega_m, h, k, \sigma_8) = (0.7, 0.3, 0.7, 1,
    0.8)$. (Others) UV luminosity functions of galaxies at redshift
    $z=7$, $3$, and $0$. In more luminous galaxies that the luminosity
    indicated by vertical dashed lines on each curve, MBHs
    successively merge. The number beside each curve shows assumed
    mass-to-light ratio to obtain each UV luminosity function.}
  \label{fig:psmf}
\end{figure*}

From the above numerical results, we conclude that the lower limit of
stellar velocity dispersion is $v_{\rm g} \sim
180~\mbox{km~s$^{-1}$}$, when the GW recoil is
$200~\mbox{km~s$^{-1}$}$. Regardless of whether the GW recoil is
exerted or not, one dominant MBH grows in model B$_0$ and B$_1$, in
both of which the stellar velocity dispersion is more than
$240~\mbox{km~s$^{-1}$}$. On the other hand, the growth of one
dominant MBH depends on the GW recoil in models C in which galaxies
have the stellar velocity dispersion of $180$~km~s$^{-1}$.  The
dependence of the MBH growth on the stellar velocity dispersion can be
understood as follows. Using Equation (\ref{eq:virial}), we express
the ratio of a Schwarzschild radius of an MBH to the virial radius of
the galaxy as
\begin{equation}
\frac{r_{{\rm sch},i}}{r_{\rm g}} = 4 \left( \frac{m_{{\rm
      B},i}}{M_{\rm g}} \right) \left( \frac{v_{\rm g}}{c}
\right)^2. \label{eq:rschoverrv}
\end{equation}
This means that the MBH horizon size is smaller compared to the galaxy
size if the stellar velocity dispersion is smaller.  Therefore, a
larger amount of angular momenta should be extracted for a binary MBH
to merger, so that the resultant largest MBH becomes less massive.

Here, we estimate the constraints for galaxy mass and size to allow
the MBH merger. In Fig.~\ref{fig:lmt}, we show the mass and size of
galaxies in which MBHs successively merge during $140 t_{\rm
  dy,g}$. They have the stellar velocity dispersion of more than
$180$~km~s$^{-1}$. Using Equation~(\ref{eq:rv}), we relate their
masses to their sizes as
\begin{equation}
  \left( \frac{M_{\rm g}}{10^{10}M_{\odot}} \right) \gtrsim
  1.5 \left( \frac{r_{\rm g}}{1\mbox{kpc}}
  \right). \label{eq:vgcnstrn}
\end{equation}
Such regions are above the solid lines in the top panels of
Fig.~\ref{fig:lmt}.

We also impose the conditions on which the successive mergers of MBHs
occurs within $1$~Gyr or within $10$~Gyr. If the merger timescale is
$140 t_{\rm dy,g}$, the galaxies should have their dynamical time of
less than $7$~Myr for $1$~Gyr case or $70$~Myr for $10$~Gyr case.
Using Equation~(\ref{eq:rv}) and (\ref{eq:tdy}), we obtain
\begin{equation}
  \left( \frac{M_{\rm g}}{10^9 M_{\odot}} \right) = 1.1 \left(
  \frac{t_{\rm dy,g}}{14 \mbox{Myr}} \right)^{-2} \left( \frac{r_{\rm
      g}}{1 \mbox{kpc}} \right)^{3}.
\end{equation}
Therefore, we can write the relation between their masses and sizes:
\begin{equation}
  \left( \frac{M_{\rm g}}{10^9~M_{\odot}} \right) > \left\{
  \begin{array}{lc}
    \displaystyle 4.4 \left( \frac{r_{\rm g}}{1\mbox{kpc}} \right)^3 &
    \cdots \mbox{$1$~Gyr case}\\ \displaystyle 0.044 \left(
    \frac{r_{\rm g}}{1\mbox{kpc}} \right)^3 & \cdots \mbox{$10$~Gyr
      case}
  \end{array}
  \right.
  \label{eq:tdycnstrn_hz}
\end{equation}
These regions are upper sides of the dashed lines in the top panels of
Fig.~\ref{fig:lmt}. As a result, the shaded regions in the top
panels of Fig.~\ref{fig:lmt} are the allowed regions for the masses
and sizes of galaxies in which MBHs can successively merge.

These constraints can be translated into those for the mass ($M_{\rm
  h}$) and size ($r_{\rm h}$) of a dark matter halo, using a
simplified model.  A dark matter halo is assumed to be six times more
massive than that of a galactic stellar component, according to the
ratio of dark matter to baryon in the universe \citep{Komatsu11}.
However, stellar components are more concentrated than the dark matter
components due to cooling when stars are formed.  Hence, the dark
matter halo does not seem to make a significant effect on the merger
dynamics.  Actually, dark matter mass at the central region is much
less than or at most comparable to stellar mass
\citep[e.g.][]{Forman85,Saglia92}. In the present analysis, the
stellar velocity dispersion is assumed to be twice of the velocity
dispersion in the dark matter halo. This is justified by the
difference between observed velocity dispersions at effective radii
and those at several effective radii in elliptical galaxies
\citep{Coccato09}. Then, the size of the dark matter halo is $24$
times larger than that of the stellar component from virial
theorem. Using these relations, we can obtain the regions of the
masses and sizes of dark matter haloes in which MBHs successively
merge, which are the shaded regions in the bottom panels of
Fig.~\ref{fig:lmt}.

The formation epoch (redshift) can be assessed depending on the masses
and sizes of dark matter haloes, in the same way as \cite{Mo98}. We
equate the size of a dark matter halo ($r_{\rm h}$) to the radius
inside which the mean mass density is $200$ times the critical density
at a given redshift $z$, and then derive the mass of a dark matter
halo ($M_{\rm h}$) inside $r_{\rm h}$. The virial mass and radius are
related as
\begin{equation}
  M_{\rm h} = 100 G^{-1} H(z)^2 r_{\rm h}^3. \label{eq:Mhalo0}
\end{equation}
We can rewrite Equation~(\ref{eq:Mhalo0}) as
\begin{equation}
  \left( \frac{M_{\rm h}}{10^{11}~M_{\odot}} \right) = 12 \left[
    \frac{H(z)}{H(10)} \right]^2 \left( \frac{r_{\rm h}}{30\mbox{kpc}}
  \right)^3. \label{eq:Mhalo1}
\end{equation}
The function $H(z)$ is expressed as
\begin{equation}
  H(z) = H_0 \bigl[ \Omega_{\rm \Lambda} + (1 - \Omega_{\rm \Lambda} -
    \Omega_{\rm m}) (1+z)^2 + \Omega_{\rm m} (1+z)^3 \bigr]^{1/2},
\end{equation}
where $H_0$ is the Hubble constant, and $\Omega_{\rm \Lambda}$ and
$\Omega_{\rm m}$ are the lambda parameter and the matter density
parameter, respectively. In Equation~(\ref{eq:Mhalo1}), we adopt
$(\Omega_{\rm \Lambda}, \Omega_{\rm m}) = (0.7, 0.3)$, and hereafter
we also adopt these values and $h=0.7$, where the Hubble constant is
$H_0 = 100h$~km~s$^{-1}$~Mpc$^{-1}$. From these equations, we can draw
the relation between masses and sizes of dark matter haloes at
formation redshifts, which is shown by dashed-dotted lines in the
bottom panels of Fig.~\ref{fig:lmt}.

From the bottom panels of Fig.~\ref{fig:lmt}, we can estimate the
minimum mass of a dark matter halo which allows MBH successive mergers
at a given redshift.  In order for the merger to occur during $1$~Gyr,
dark matter haloes formed at redshift $z=7$ should have more than $4
\times 10^{10} M_{\odot}$, which corresponds to the stellar component
mass of about $6.7 \times 10^{9} M_{\odot}$.  In the case of mergers
during $10$~Gyr, dark matter haloes formed at redshift $z=3$ should
have more than $7 \times 10^{10} M_{\odot}$.

If there were only one binary MBH in a nonaxisymmetric galactic
potential, the timescale for the merger is about $10$~Gyr or $0.3$~Gyr
respectively for the stellar component of $10^9M_{\odot}$ or
$10^{11}M_{\odot}$ \citep{Khan11}.  Therefore, at redshifts of $z
\gtrsim 7$, when the cosmic age is less than $1$~Gyr, another MBH may
intrude before a binary MBH merges.  Our results show that even if
multiple MBHs exist in a galaxy with $4 \times 10^{10} M_{\odot}$ at
redshift $z=7$, the MBHs can successively merge.

\subsection{Merger criteria for galactic luminosity}

In the above, we have derived the constraints for the masses and sizes
of galaxies and their parent dark haloes, in which the successive
merger of MBHs can occur.  Here, we compare the luminosity function
based on the Press-Schechter formalism. The luminosity function of
galaxies is obtained as follows.  We can give an ultraviolet (UV)
magnitude of a galaxy embedded in a halo with mass $M_{\rm h}$ as
\begin{equation}
  M_{\rm UV} = M_{{\rm UV},\odot} - \frac{2.5}{\log{10}} \log \left[
    \left( \frac{M_{\rm h}}{M_{\odot}} \right) \left(
    \frac{\Omega_{\rm b}}{\Omega_{\rm m}} \right) \Upsilon_{{\rm
        UV},\odot}^{-1} \right],
\end{equation}
where $M_{{\rm UV},\odot}$ is UV magnitude of the Sun, $\Upsilon_{{\rm
    UV},\odot}$ is the mass-to-UV luminosity ratios scaled by that of
the Sun. We set $M_{{\rm UV},\odot}=5.6$.  Here, we assume that a
galaxy mass $M_{\rm g}$ is equal to $(\Omega_{\rm m}/\Omega_{\rm
  b})^{-1}M_{\rm h}$, and that $\Omega_{\rm m}/\Omega_{\rm b}=6$. We
denote the number density of haloes by $n$. Note that $n$ can be
regarded as the number density of galaxies, since we assume a halo has
one galaxy. Then, Press-Schechter mass function of dark matter haloes
(shown in the top left panel of Fig.~\ref{fig:psmf}) can be
transformed into the luminosity function at UV band as
\begin{align} 
  \frac{dn}{dM_{\rm UV}} &= - \frac{d\left(M_{\rm
      h}/M_{\odot}\right)}{dM_{\rm UV}} \frac{dn}{d(M_{\rm
      h}/M_{\odot})} \\ &= \frac{\log{10}}{2.5} \left( \frac{M_{\rm
      h}}{M_{\odot}} \right) \frac{dn}{d(M_{\rm
      h}/M_{\odot})}.
\end{align}

Observationally, the mass-to-luminosity ratios for high-redshift Ly
$\alpha$ emitters (LAEs), $\Upsilon_{{\rm UV},\odot}$, can range from
0.3 to 10 \citep{Fernandez08}.  For low-redshift galaxies, the
mass-to-luminosity ratios range from 2 to 10 in normal galaxies, but
reach $\sim 100$ in dwarf galaxies \citep{Hirashita98,Strigari08}.
Considering the observed mass-to-luminosity ratios, we draw the UV
luminosity function $dn/d(M_{\rm UV}/M_{\odot})$ in
Fig.~\ref{fig:psmf} for the cases of redshift $z=7$ (top right), $3$
(bottom left), and $0$ (bottom right). In each panel, we show the UV
luminosity function $dn/d(M_{\rm UV}/M_{\odot})$ with different
$\Upsilon_{{\rm UV},\odot}$. The values of $\Upsilon_{{\rm UV},\odot}$
are indicated by numbers beside the curves. The critical luminosity of
galaxies for the successive mergers is shown by vertical dashed lines
attached with each curve. The successive mergers happen in galaxies
brighter than the critical luminosities.

Observed luminosity functions of high-redshift LAEs \citep{Ouchi09}
and those of Lyman break galaxies (LBGs) \citep{Jiang11} seem to match
well the curves with $\Upsilon_{{\rm UV},\odot} \sim 1$.  Thus, the
successive MBH merger is expected for LAEs or LBGs brighter than
$M_{\rm UV}\simeq - 19$ (see the top right and bottom left panels of
Fig.~\ref{fig:psmf}). Since $\Upsilon_{{\rm UV},\odot} = 2$ -- $10$ in
low-redshift galaxies, the successive merger is expected for
low-redshift galaxies brighter than $M_{\rm UV}\simeq - 18$ (see the
bottom right panel of Fig.~\ref{fig:psmf}). Note that if
$\Upsilon_{{\rm UV},\odot}$ is as large as 100 in dwarf galaxies, the
critical UV magnitude can be overestimated by about a factor of
two. These dwarf galaxies would lose an amount of baryons through
their evolution. Although we adopt $\Omega_{\rm m}/\Omega_{\rm b}=6$
for all galaxies, $\Omega_{\rm m}/\Omega_{\rm b}$ should be set to a
larger value for these dwarf galaxies. If we do so, the curve of
$\Upsilon_{{\rm UV},\odot}=100$ in the bottom right panel of
Fig.~\ref{fig:psmf} will shift leftward.

\section{Back-reaction to a Host Galaxy}
\label{sec:back-reaction}

\subsection{Galaxy structure}

\begin{figure}
  \begin{center}
    \includegraphics[scale=1.4]{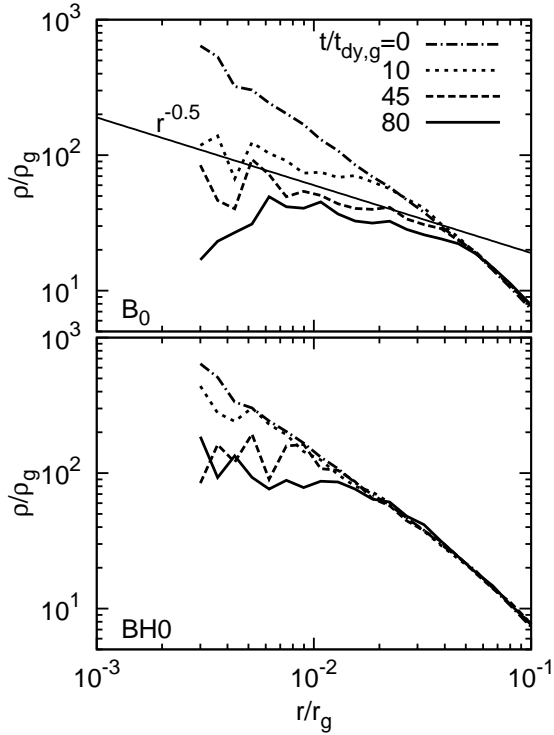}
  \end{center}
  \caption{Mass density profile of stars in a galaxy with ten MBHs
    (top) or without MBHs (bottom). In both panels, dashed-dotted,
    dotted, dashed, and solid curves indicate the mass density at
    $t/t_{\rm dy,g}=0$, $10$, $45$, and $80$, respectively. A solid
    line in the top panel shows the relation of $\rho \propto
    r^{-0.5}$.}
  \label{fig:evl_rho}
\end{figure}

\begin{figure}
  \begin{center}
    \includegraphics[scale=1.5]{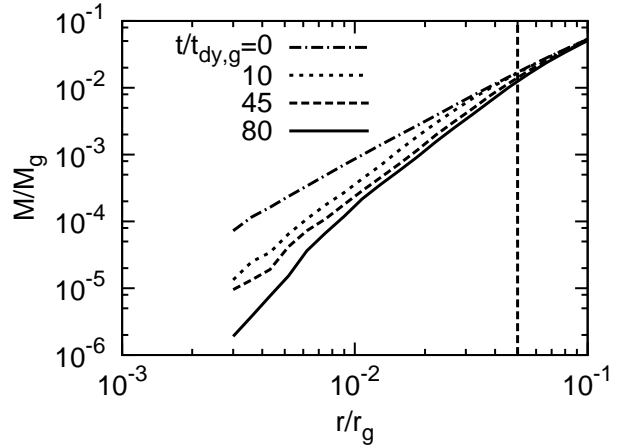}
  \end{center}
  \caption{Mass of stars within each radius in model B$_0$ with ten
    MBHs. Dashed-dotted, dotted, dashed, and solid curves indicate the
    mass at the time $t/t_{\rm dy,g}=0, 10, 45$, and $80$. The
    vertical dashed line indicates $r/r_{\rm g}=0.05$.}
  \label{fig:evl_mr}
\end{figure}

Hereafter, we focus on the simulation results of model B$_0$. If
necessary, we can compare a simulation including the GW recoil, model
B$_1$. Fig.~\ref{fig:evl_rho} shows the evolution of mass density
profile of stars (the top panel). The mass density inside $r/r_{\rm g}
= 0.05$ decreases gradually. This is because MBHs give their kinetic
energy to stars as a back reaction of dynamical friction and
sling-shot mechanism.  Until $t/t_{\rm dy,g}=80$, six MBHs merge. We
can see in the top panel of Fig.~\ref{fig:evl_rho} that the mass
density profile is roughly proportional to $r^{-0.5}$ in the range
from $r/r_{\rm g}=5\times 10^{-3}$ to $r/r_{\rm g}=0.05$. Such a
density slope is consistent with those in a galaxy with two MBHs and
three MBHs \citep{Nakano99,Iwasawa08}.  In Fig.~\ref{fig:evl_mr}, we
see an enclosed mass of the galaxy within $r/r_{\rm g}=0.05$ (vertical
dashed line) is $10^{-2}M_{\rm g}$, which is ten times higher than the
total mass of MBHs.  Hence, the present simulation shows that MBHs can
affect the galactic structure of the central regions that include
about ten times the total mass of MBHs.

\begin{figure}
  \begin{center}
    \includegraphics[scale=1.5]{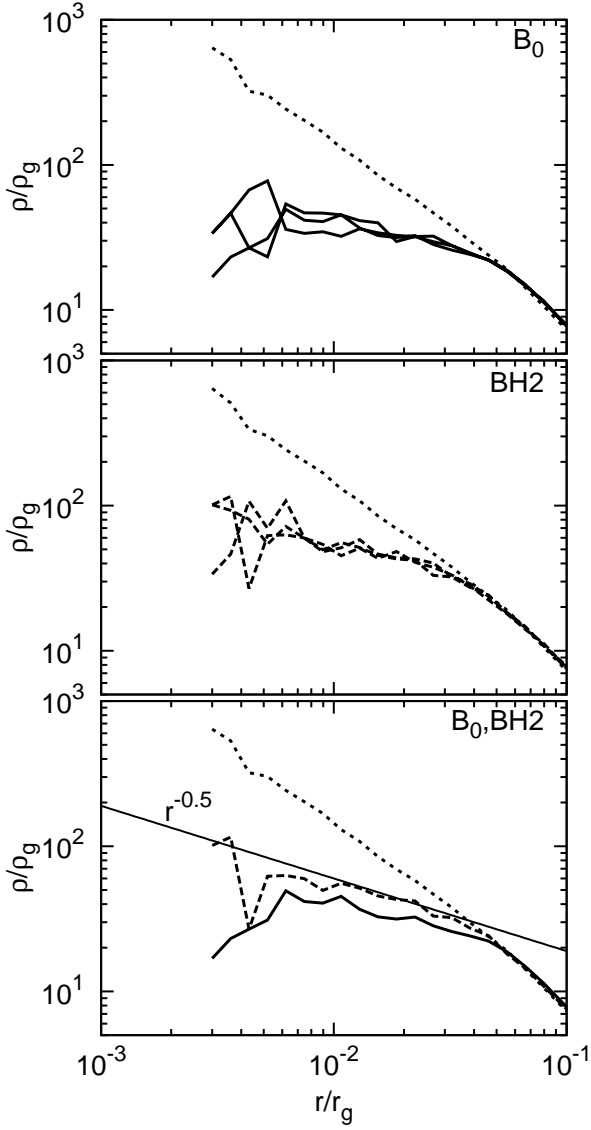}
  \end{center}
  \caption{Mass density profile of a galaxy with ten MBHs (top), that
    with two MBHs (middle), and both (bottom). In all the panels, the
    dotted curve shows the profile at the initial time. In the top
    panel, the solid curves indicate the profile at the time $t/t_{\rm
      dy,g}=79, 80$, and $81$. In the middle panel, the dashed curves
    indicate the profile at the time $t/t_{\rm dy,g}=30, 40$, and
    $50$. In the bottom panel, the solid and dashed curves are the
    profiles of a galaxy with ten MBHs at $t/t_{\rm dy,g}=80$, and a
    galaxy with two MBHs at $t/t_{\rm dy,g}=50$, respectively.}
  \label{fig:rho}
\end{figure}

We also compare the structure of the galaxy containing ten MBHs to
that containing two MBHs. The total masses of MBHs are the same, that
is, $0.1$ \% of the galaxy mass in both of the models. The top panel
of Fig.~\ref{fig:rho} shows the mass density profile of the galaxy
with ten MBHs at $t/t_{\rm dy,g}=79, 80$ and $81$. The mass density is
not fluctuated on the dynamical timescale in the range from $r/r_{\rm
  g}=0.01$ to $0.05$. In the middle panel of Fig.~\ref{fig:rho}, we
show the mass density profile of the galaxy with two MBHs at $t/t_{\rm
  dy,g}=30, 40$, and $50$. During $20$ dynamical time, the mass
density profile is not changed in the case of the galaxy with two
MBHs. We expect that the mass density profile is never changed after
$t/t_{\rm dy,g}=50$.

In the bottom panel of Fig.~\ref{fig:rho}, we compare the mass density
profile of the galaxy containing ten MBHs at $t/t_{\rm dy,g}=80$ with
those containing two MBHs at $t/t_{\rm dy,g}=50$. Both of the mass
density slopes are proportional to $r^{-0.5}$. However, the mass
density of the galaxy with ten MBHs is lower by a factor of $1.5$ than
that with two MBHs. This difference results from the sling-shot
mechanism in the galaxy with ten MBHs. Owing to the sling-shot
mechanism among three MBHs, MBHs receive kinetic energy, and transfer
their kinetic energy to stars through the dynamical friction. Such
picture is consistent with a galaxy with three MBHs \citep{Iwasawa08}.

Here, we verify that the central density of a galaxy in model B$_0$
are decreased by MBH dynamics, not by artificial two-body relaxation.
The bottom panel of Fig.~\ref{fig:evl_rho} shows the evolution of mass
density profile of a galaxy without any MBH, in which only two-body
relaxation decreases the central mass density of the galaxy. Comparing
mass densities in the top and bottom panels of Fig.~\ref{fig:evl_rho},
we can see that the central density in model B$_0$ is decreased much
more rapidly than that in model BH0. Therefore, the central density of
a galaxy in model B$_0$ are dominantly decreased by MBH dynamics.

\subsection{High-velocity stars}

We investigate stars which are ejected from a galaxy with high
speed. Such stars are generated through the sling-shot mechanism
induced by a binary MBH. We focus on the simulation results of model
B$_0$. If necessary, we can compare a simulation including the GW
recoil, model B$_1$, in which MBHs also successively merge.

\begin{figure}
  \begin{center}
    \includegraphics[scale=1.0]{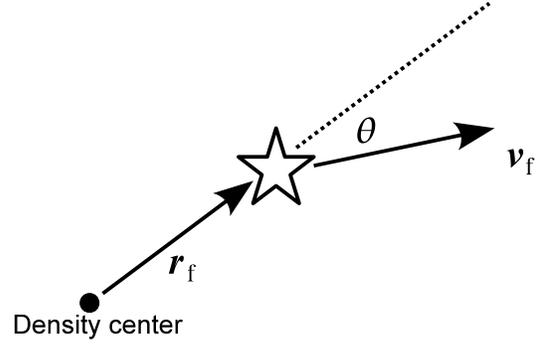}
  \end{center}
  \caption{Illustration of the position vector ($\bm{r}_{\rm f}$),
    velocity vector ($\bm{v}_{\rm f}$) of a star, and $\theta$.}
  \label{fig:theta}
\end{figure}

We compare the velocity distributions of stars as a function of
$\theta$ at the time $t/t_{\rm dy,g}=0$ and $80$ in the cases of
models with and without MBHs.  We illustrate the relation of $\theta$
to the position and velocity vectors in Fig.~\ref{fig:theta}.  Then,
the $\theta$ is expressed as
\begin{equation}
  \theta = \cos^{-1}\left(\frac{\bm{r}_{\rm f} \cdot
    \bm{v}_{\rm f}}{r_{\rm f} v_{\rm f}}\right),
\end{equation}
where $\bm{r}_{\rm f}$ and $\bm{v}_{\rm f}$ are respectively
the position and velocity vectors of a star, and $r_{\rm
  f}=|\bm{r}_{\rm f}|$ and $v_{\rm f}=|\bm{v}_{\rm f}|$. The
origin of the position vector is set to the galaxy centre.

\begin{figure}
  \begin{center}
    \includegraphics[scale=1.2]{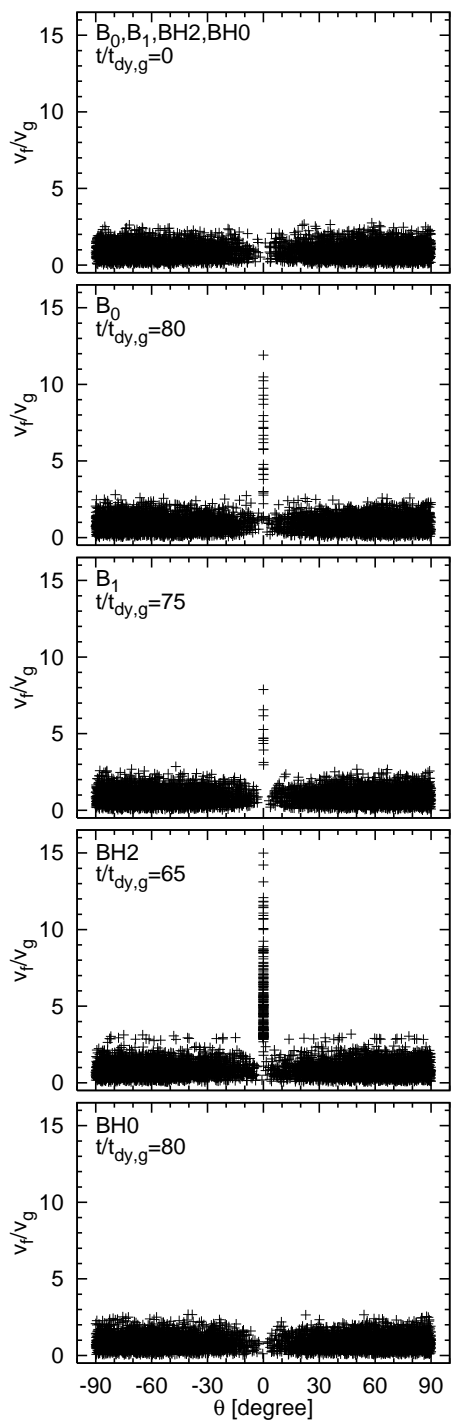}
  \end{center}
  \caption{Velocity distributions of stars as a function of $\theta$
    in models B$_0$, B$_1$, BH2, and BH0 at the initial time (top), in
    model B$_0$ at $t/t_{\rm dy,g}=80$ (second top), in model B$_1$ at
    $t/t_{\rm dy,g}=75$ (middle), in model BH2 at $t/t_{\rm dy,g}=65$
    (second bottom), and in model BH0 at $t/t_{\rm dy,g}=80$
    (bottom).}
  \label{fig:ang_hvs}
\end{figure}

We define high-velocity stars as stars whose velocities are more than
$2\sqrt{2}v_{\rm g}$. Fig.~\ref{fig:ang_hvs} shows the resultant
velocity distributions of stars. The presence of high-velocity stars
is an outstanding feature of model B$_0$ at the time $t_{\rm dy,g}=80$
(the second top panel). We also find such high-velocity stars in a
galaxy with two MBHs, model BH2 (the second bottom panel). We can see
that some stars have velocities higher than $v_{\rm f}/v_{\rm
  g}=10$. Furthermore, they have extremely radial orbits around
$\theta=0$.  This is because they are generated at the galactic centre
through the sling-shot mechanism by a binary MBH, and directly go away
outside the galaxy. Note that no high-velocity star is generated in a
galaxy without MBHs, model BH0 (see the bottom panel of
Fig.~\ref{fig:ang_hvs}).

\begin{figure}
  \begin{center}
    \includegraphics[scale=1.5]{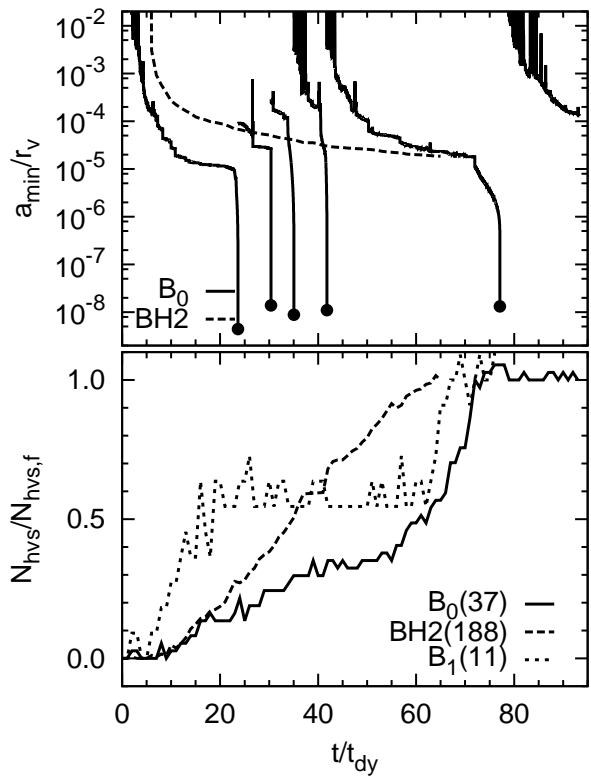}
  \end{center}
  \caption{Time evolution of the minimum semi-major axis of a binary
    MBH at each time (top), and the number of high-velocity stars
    scaled by its final number (bottom). High-velocity stars are
    defined as those whose velocities are more than $2\sqrt{2}v_{\rm
      g}$.}
  \label{fig:cfunc_hvs}
\end{figure}

We investigate the difference between properties of high-velocity
stars in the cases of galaxies with ten MBHs and with two MBHs. $37$
high-velocity stars have been generated at $t/t_{\rm dy,g}=80$ in
model B$_0$, in contrast to $188$ high-velocity stars at $t/t_{\rm
  dy,g}=50$ in model BH2. The generation rate of high-velocity stars
in model B$_0$ is ten times lower than that in model BH2. This is
because a galaxy with ten MBHs does not always have a binary MBH with
small semi-major axis, while a binary MBH stays in the central region
of the galaxy in model BH2 (see the top panel of
Fig.~\ref{fig:cfunc_hvs}). It is difficult for a binary MBH to produce
high-velocity stars, unless its semi-major axis is as small as $\sim
10^{-5} r_{\rm g}$. This is readily estimated with numerical results
by \cite{Quinlan96}. A binary MBH typically gives a kick velocity of
the order of $v_{\rm kick} = \sqrt{2.5G\mu/a}$ to a field star through
the sling-shot mechanism, where $\mu$ and $a$ are the reduced mass and
semi-major axis of a binary MBH. Supposing that a binary MBH consists
of two MBHs with the initial mass, a binary with the semi-major axis
of $a<3.3 \times 10^{-5} r_{\rm g}$ can produce high-velocity stars
with $v_{\rm kick}>2\sqrt{2}v_{\rm g}$.

Another possible reason is that more stars interact with a binary MBH
in model BH2, since the total mass of the binary MBH in model BH2 is
larger than that in model B$_0$.

The total number of high-velocity stars increases in different ways
between models B$_0$ and BH2. As seen in the bottom panel of
Fig.~\ref{fig:cfunc_hvs}, the number of high-velocity stars increases
at a roughly constant rate in model BH2 from the time $t/t_{\rm
  dy,g}=10$ to $50$. On the other hand, the generation rate of
high-velocity stars is largely changed in model B$_0$ from the time
$t/t_{\rm dy,g}=0$ to $80$ (see the solid curve in the bottom panel of
Fig.~\ref{fig:cfunc_hvs}). The generation rate is low during the time
$t/t_{\rm dy,g}=40$ -- $60$, and during $t/t_{\rm dy,g}=70$ -- $90$,
while it is high during $t/t_{\rm dy,g}=10$ -- $40$ and during
$t/t_{\rm dy,g}=60$ -- $70$. This feature is similar to high-velocity
stars in model B$_1$.

In a galaxy with ten MBHs, such as models B$_0$ and B$_1$,
high-velocity stars are generated intermittently because of the
occasional absence of a binary MBH whose semi-major axis is favourable
to eject stars, $\sim 10^{-5} r_{\rm g}$. This can be verified in
models B$_0$ and BH2. During $t_{\rm dy,g}=10$ -- $40$, and $60$ --
$70$, high-velocity stars are generated at a high rate in model B$_0$
(see the bottom panel of Fig.~\ref{fig:cfunc_hvs}). At this time,
there is a binary MBH with semi-major axis of about $10^{-5}r_{\rm g}$
(see the top panel of Fig.~\ref{fig:cfunc_hvs}). The generation rate
of high-velocity stars is low during $t_{\rm dy,g}=10$ -- $20$, $40$
-- $60$, and $70$ -- $90$. Except $t_{\rm dy,g}=70$ -- $80$, there is
a binary MBH with semi-major axis much larger than $10^{-5} r_{\rm
  g}$, or no binary MBH.  Therefore, stars are not ejected through
sling-shot mechanism. During $t_{\rm dy,g}=70$ -- $80$, there is a
binary MBH with semi-major axis much less than $10^{-5} r_{\rm
  g}$. Such a binary MBH cannot interact with stars due to small cross
section. On the other hand, there is a binary MBH with semi-major axis
$\sim 10^{-5} r_{\rm g}$ in model BH2.

The feature of the intermittent generation rate can be a useful probe
to constrain the formation mechanism of a single merged MBH, or a
binary MBH at the galaxy centre at the present time.

\section{Summary}
\label{sec:summary}

We have performed $N$-body simulations to investigate successive
mergers of MBHs in galaxies with different masses and radii. We have
found that about a half of multiple MBHs successively merge to one
bigger MBH within $140t_{\rm dy,g}$ in galaxies with the velocity
dispersion larger than $\sim 180$~km~s$^{-1}$. The merger of MBHs is
promoted, such that the loss cone of binary MBHs is refilled by MBHs
losing their angular momenta due to dynamical friction. GW recoil does
not affect the merger process, if the recoil velocity is of the order
of the stellar velocity dispersion.  Galaxies which allow multiple
MBHs to merge should reside in dark matter haloes with the mass more
than $4 \times 10^{10}M_{\odot}$, if these dark matter haloes form at
high redshifts. These galaxies could correspond to LAEs or LBGs
brighter than the UV magnitude $M_{\rm UV} \simeq -19$ at high
redshifts.  On the other hand, an MBH which has experienced the
successive merger can inhabit low-redshift galaxies brighter than
$M_{\rm UV} \simeq -18$.

We have also investigated the evolution of the galactic structure and
the generation of high-velocity stars as the back-reaction by the
successive merger of MBHs.  We have found that the dynamics of MBHs
affects the central regions of galaxy that contain about ten times the
total mass of MBHs. The mass density profile is transformed to $\rho
\propto r^{-0.5}$, which is the same as the mass density profile in
the case of a galaxy with two and three MBHs. The mass density in the
central regions is $1.5$ times smaller than in the case of the galaxy
with two MBHs.  In a galaxy with ten MBHs, high-velocity stars are
generated intermittently, while they are generated at a constant rate
in the case of a galaxy with two MBHs. Such features should enable us
to constrain the merger mechanism of MBHs in a galaxy.

\section*{Acknowledgements}

We thank Kohji Yoshikawa for fruitful advice on cosmological
arguments, Masaki Iwasawa for useful comments about our simulation
method, and Yuichi Matsuda for stimulating discussion.
Numerical simulations have been performed with computational
facilities at the Center for Computational Sciences in the University
of Tsukuba. This work was supported in part by the FIRST project based
on the Grants-in-Aid for Specially Promoted Research by MEXT
(16002003), and Grant-in-Aid for Scientific Research (S) by JSPS
(20224002).

\bsp

\label{lastpage}

\end{document}